\documentclass[prl,twocolumn,amsmath,amssymb,floatfix,superscriptaddress,nofootinbib,preprintnumbers]{revtex4-2}
\pdfoutput=1

\usepackage{amssymb}
\usepackage{amsmath}

\usepackage{graphicx}
\usepackage{graphics}
\usepackage{dcolumn}
\usepackage[dvipsnames]{xcolor}
\usepackage{fancyhdr} 
\usepackage{graphicx}
\usepackage{float}
\usepackage{multirow}

\usepackage{subfig}
\usepackage[colorlinks=true,linktocpage=true,urlcolor=blue,citecolor=blue,linkcolor=blue]{hyperref}

\usepackage[utf8]{inputenc}

\usepackage{dsfont}

\usepackage[justification=centerlast, format=plain, labelfont=bf]{caption}

\makeatletter
\renewcommand\tableofcontents{%
    \@starttoc{toc}%
}
\makeatother

\DeclareCaptionJustification{justified}{\leftskip=0pt \rightskip=0pt \parfillskip=0pt plus 1fil}

    \newcommand{\be}{\begin{equation}}
  \newcommand{\ee}{\end{equation}}
    \newcommand{\ba}{\begin{align}}
  \newcommand{\ea}{\end{align}}

\newcommand{\Msun}{M_{\odot}}
\newcommand{\Mpcinv}{ {\rm Mpc}^{-1} }

\newcommand{\MUV}{ M_{\rm UV} }

\definecolor{atomictangerine}{rgb}{1.0, 0.6, 0.4}

\makeatletter
\def\doauthor#1#2#3{%
  \ignorespaces#1\unskip
  \begingroup
   #3%
  \@if@empty{#2}{\@listcomma\endgroup{}{}}{\endgroup{\comma@space}{}\frontmatter@footnote{#2}}%
  \space \@listand
}%
\makeatother

\makeatletter
\def\@ssect@ltx#1#2#3#4#5#6[#7]#8{%
  \def\H@svsec{\phantomsection}%
  \@tempskipa #5\relax
  \@ifdim{\@tempskipa>\z@}{%
    \begingroup
      \interlinepenalty \@M
      #6{%
       \@ifundefined{@hangfroms@#1}{\@hang@froms}{\csname @hangfroms@#1\endcsname}%
       {\hskip#3\relax\H@svsec}{#8}%
      }%
      \@@par
    \endgroup
    \@ifundefined{#1smark}{\@gobble}{\csname #1smark\endcsname}{#7}%
  }{%
    \def\@svsechd{%
      #6{%
       \@ifundefined{@runin@tos@#1}{\@runin@tos}{\csname @runin@tos@#1\endcsname}%
       {\hskip#3\relax\H@svsec}{#8}%
      }%
      \@ifundefined{#1smark}{\@gobble}{\csname #1smark\endcsname}{#7}%
      \addcontentsline{toc}{#1}{\protect\numberline{}#8}%
    }%
  }%
  \@xsect{#5}%
}%
\makeatother

\begin{document}

\preprint{KCL-2021-75}

\title{\vspace{-0.23cm}New Roads to the Small-Scale Universe: Measurements of the Clustering of Matter with the High-Redshift UV Galaxy Luminosity Function}

\author{Nashwan Sabti$^{\mathds{S},}$}
\affiliation{Department of Physics, King's College London, Strand, London WC2R 2LS, UK}
\author{Julian B. Mu\~{n}oz$^{\mathds{M},}$}
\affiliation{Harvard-Smithsonian Center for Astrophysics, Cambridge, MA 02138, USA}
\author{Diego Blas$^{\mathds{B},}$}
\affiliation{Grup de F\'isica Te\`orica, 
Departament  de  F\'isica, Universitat  Aut\`onoma  de  Barcelona,   Bellaterra, 08193 Barcelona, Spain}
\affiliation{Institut de Fisica d’Altes Energies (IFAE), The Barcelona Institute of Science and Technology,\\ Campus UAB, 08193 Bellaterra  (Barcelona), Spain}

\def\thefootnote{$\mathds{S}$\hspace{0.7pt}}\footnotetext{\href{mailto:nashwan.sabti@kcl.ac.uk}{nashwan.sabti@kcl.ac.uk}}
\def\thefootnote{$\mathds{M}$\hspace{-0.9pt}}\footnotetext{\href{mailto:julianmunoz@cfa.harvard.edu}{julianmunoz@cfa.harvard.edu}}
\def\thefootnote{$\mathds{B}$}\footnotetext{\href{mailto:dblas@ifae.es}{dblas@ifae.es}}
\setcounter{footnote}{0}
\def\thefootnote{\arabic{footnote}}

\begin{abstract}
    \noindent The epochs of cosmic dawn and reionisation present promising avenues for understanding the role of dark matter (DM) in our cosmos. The first galaxies that populated the Universe during these eras resided in DM halos that were much less massive than their counterparts today. Consequently, observations of such galaxies can provide us with a handle on the clustering of DM in an otherwise currently inaccessible regime. In this work, we use high-redshift UV galaxy luminosity-function (UV LF) data from the Hubble Space Telescope to study the clustering properties of DM at small scales. In particular, we present new measurements of the matter power spectrum at wavenumbers $0.5\,\mathrm{Mpc}^{-1} < k < 10\,\mathrm{Mpc}^{-1}$ to roughly 30\% precision, obtained after marginalising over the unknown astrophysics. These new data points cover the uncharted redshift range $4\leq z\leq 10$ and encompass scales beyond those probed by Cosmic-Microwave-Background and large-scale-structure observations. This work establishes the UV LF as a powerful tool to probe the nature of DM in a different regime than other cosmological and astrophysical data sets.
\end{abstract}


\maketitle
\enlargethispage{1.1cm}

\textit{Introduction.\ ---} Our exploration of the Universe has entered an era where its fundamental properties can be studied with multiple probes in a complementary way. This has allowed us to track its evolution not only from the time of primordial nucleosynthesis down to the present day, but also across length scales that span several orders of magnitude~\cite{Chabanier:2019eai}. Observations show us a rich evolution of cosmic structures that started off as tiny fluctuations at the time of photon decoupling, and hierarchically grew to become the cosmological large-scale structure (LSS) today~\cite{Frenk:2012ph}. These measurements have been exploited to learn about the mechanisms underlying the formation and growth of structure~\cite{Croft:2000hs, Allen:2011zs, Kilbinger:2014cea, Akrami:2018vks}, pointing towards a consistent framework that describes the data in the observed range.

There are, however, still outstanding questions and challenges to our understanding of structure formation~\cite{Bull:2015stt}. 
Chief among them is the nature of dark matter (DM), which along with baryons forms the LSS of our Universe. Current data suggest that DM is cold and collisionless at super-galactic scales~\cite{Blumenthal:1984bp,Bertone:2016nfn}, while at smaller scales the situation is more uncertain. Alternatives to the cold-DM (CDM) paradigm typically show a different behaviour at these small scales, making this an interesting regime to probe the properties of DM~\cite{Weinberg:2013aya, DelPopolo:2016emo,Bullock:2017xww, deMartino:2020gfi}. 
Another example comes from measurements of the large-scale clustering amplitude $\sigma_8$ using observations of the (high-redshift) Cosmic Microwave Background (CMB) and the (low-redshift) LSS, which appear to be in slight tension with each other~\cite{Verde:2019ivm,DiValentino:2020vvd, Heymans:2020gsg, Perivolaropoulos:2021jda}.


A promising probe to tackle these open questions is the UV galaxy luminosity function (UV LF). The UV LF captures the abundance of galaxies as a function of their magnitude (or, equivalently, luminosity) at different points in the cosmic history and, therefore, contains a wealth of information on the physics of galaxy formation. The past decade has seen the establishment of UV LF catalogues that cover tens of thousands of galaxies during the (pre-)reionisation era~\cite{Bouwens:2014fua,Finkelstein_2015,Atek:2015axa,Livermore:2016mbs,Bouwens_2017asdasd,Mehta_2017,Ishigaki_2018,Oesch_2018,Atek:2018nsc, Rojas_Ruiz_2020, Bouwens_2021}. By using the abundance of these galaxies as an indirect probe of the mass function of DM halos, we now have a new handle on the physics of structure formation in this uncharted epoch.

\begin{figure}[t!]
    \centering
    \includegraphics[width=\linewidth]{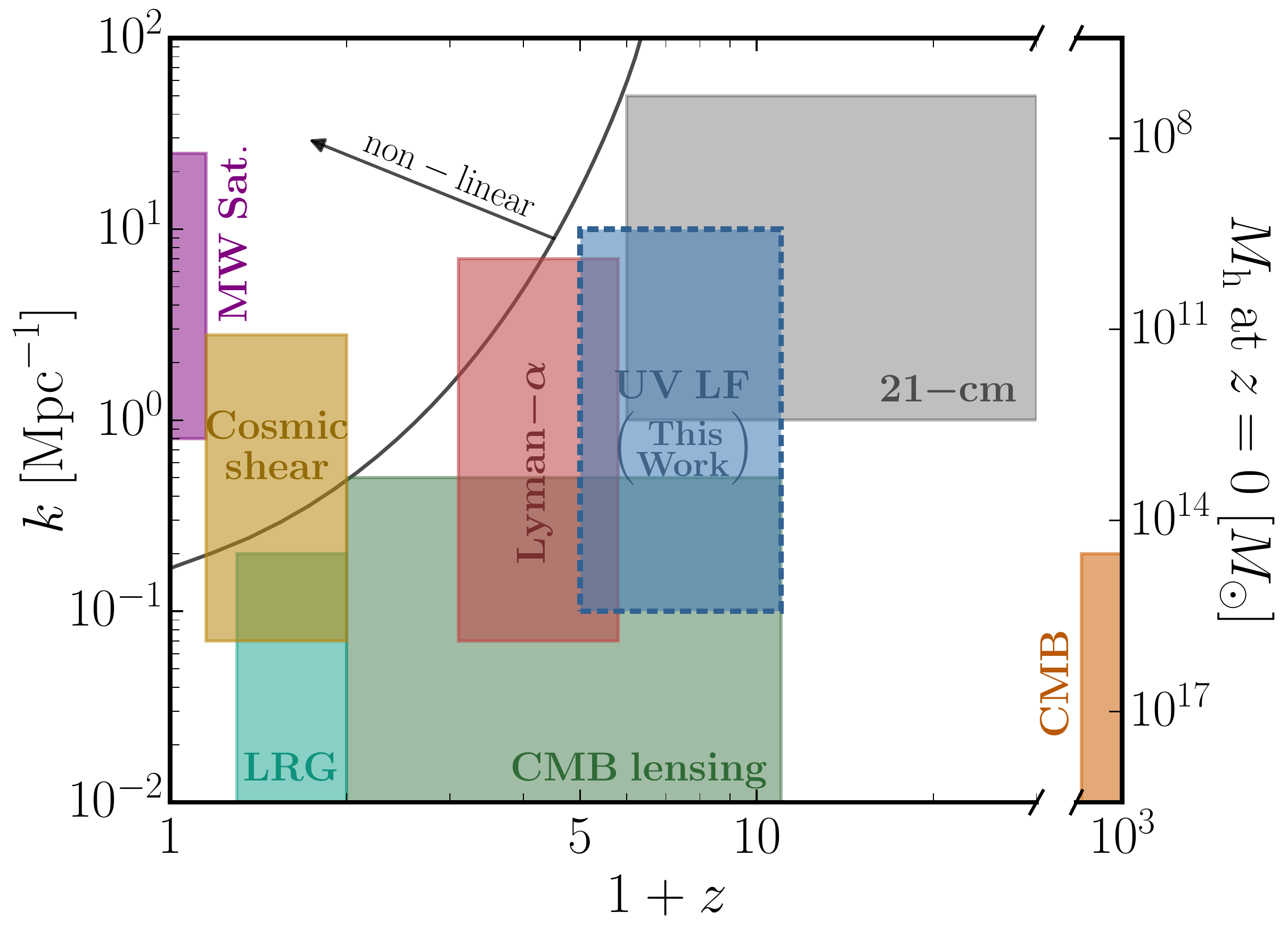}
    \caption{Illustration of the redshift and wavenumber ranges probed by different types of observations. These include Milky-Way (MW) satellites~\cite{Banik:2019smi}, cosmic-shear~\cite{DES:2021wwk} and Luminous Red Galaxies (LRG)~\cite{Chabanier:2019eai} surveys, CMB~\cite{Planck:2018nkj} and CMB lensing~\cite{CMB-S4:2016ple} observations, the Lyman-$\alpha$ forest~\cite{Chabanier:2019eai}, and (future) 21-cm~\cite{Munoz:2019hjh} data. The blue region corresponds to our UV LF studies, and covers scales and times that are currently inaccessible with other probes. For reference, the right axis is a rough estimate of the corresponding halo masses at redshift $z=0$, and the region above the black line indicates the non-linear regime.}
    \label{fig:big_picture}
\end{figure}

In this {\it Letter}, we make use of UV galaxy luminosity function data to measure the clustering of matter at small scales ($k\sim 0.5-10\, \Mpcinv$) and high redshifts ($z=4-10$). We show the reach of the UV LF in Fig.~\ref{fig:big_picture}, where it is clear that it presents a unique opportunity to study the state of the Universe in a complementary range to local-Universe probes and CMB observations. The key element of our work is a robust analysis pipeline that carefully marginalises over astrophysical uncertainties, including the parameters that enter the halo-galaxy connection. Specifically, we use the publicly available likelihood code \texttt{GALLUMI}\footnote{\vspace{-0.85cm}\href{https://github.com/NNSSA/GALLUMI_public}{https://github.com/NNSSA/GALLUMI\_public}}, which we introduce in our companion paper~\cite{Sabti:2021xvh}, to perform our analysis. \texttt{GALLUMI} is implemented in the MCMC sampler \texttt{MontePython}~\cite{Audren:2012wb,Brinckmann:2018cvx} and can be readily run in conjunction with other data sets. We find that our determination of the matter power spectrum is in agreement with the standard $\Lambda$CDM prediction over the entire range of wavenumbers studied here, with an accuracy down to a few tens of percents.

\newpage

\textit{UV LF Data.\ ---} We use galaxy abundance measurements gathered over the last decade with the Hubble Space Telescope (HST). In particular, we will perform our analysis with the data from~\cite{Oesch_2018, Bouwens_2021}, which compile search results from the Hubble Legacy Fields and Frontier Fields programs to determine the UV LF over the redshift range $z = 4-10$. These data are based on blank- and parallel-field observations, where galaxies were selected using an object classifier~\cite{Bertin:1996fj},
alongside color criteria requirements in selection techniques similar to the Lyman-break dropout method~\cite{Steidel:1996ym}. Importantly, galaxies behind lensing clusters were excluded to avoid systematic errors that may arise during the construction of lensing models~\cite{Bouwens_2017asdasd}. The data as is does not account for cosmic variance, nor the attenuation caused by dust extinction. In addition, since the UV LF is defined in terms of number densities, the data is reported within a certain fiducial cosmology (a flat $\Lambda$CDM universe\footnote{Throughout this work, we will fix the total sum of neutrino masses to $\sum m_\nu = 0.06\,\mathrm{eV}$. This choice allows us to make comparisons with other analyses, but has negligible impact on our results.}, with Hubble parameter $H_0 = 70\,\mathrm{km}\,\mathrm{s}^{-1}\mathrm{Mpc}^{-1}$ and matter density parameter $\Omega_\mathrm{m} = 0.3$). We correct the UV LF for all three points using the methods described in detail in our companion paper~\cite{Sabti:2021xvh}. In short, we correct for the Alcock-Paczy\'{n}ski effect~\cite{Alcock:1979mp}, use the IRX$-\beta$ relationship~\cite{Meurer:1999jj} with the calibration from~\cite{Overzier:2010aa} to compute the dust attenuation, and impose a minimal error of 20\% on each individual data point to account for cosmic variance.\\

\textit{Formalism and Models.\ ---} In order to translate the UV emission and abundance of high-redshift galaxies to cosmological parameters, we need to consider two separate components. 
The UV LF is defined as:
\begin{align}
    \label{eq:UVLFdef}
    \Phi_\mathrm{UV} = \dfrac{\mathrm{d}n}{\mathrm{d}M_\mathrm{h}} \times \dfrac{\mathrm{d}M_\mathrm{h}}{\mathrm{d} \MUV}\ ,
\end{align}
where the first term (the halo mass function, HMF) mainly depends on cosmology, whereas the second term (the halo-galaxy connection, which links the mass $M_\mathrm{h}$ of a DM halo to the absolute magnitude $M_\mathrm{UV}$ of the galaxy it hosts) depends on astrophysics. Here we implicitly assumed that the halo occupation distribution is unity, i.e., each halo hosts one central galaxy, which is a good approximation at these high redshifts~\cite{Bhowmick:2018waq}. 

For the HMF, we make use of the Sheth-Tormen mass function, given by~\cite{Sheth:2001dp}:
\begin{align}
    \label{eq:HMF}
    \frac{\mathrm{d}n_\mathrm{h}}{\mathrm{d}M_\mathrm{h}} = \frac{\overline{\rho}_\mathrm{m}}{M_\mathrm{h}}\frac{\mathrm{d}\ln\sigma_{M_\mathrm{h}}^{-1}}{\mathrm{d}M_\mathrm{h}}f_\mathrm{ST}(\sigma_{M_\mathrm{h}})\ ,
\end{align}
with
\begin{align}
    \label{eq:f_ST}
    f_\mathrm{ST}(\sigma_{M_\mathrm{h}}) =\ & A_\mathrm{ST}\sqrt{\frac{2a_\mathrm{ST}}{\pi}}\left[1+\left(\frac{\sigma_{M_\mathrm{h}}^2}{a_\mathrm{ST}\delta_\mathrm{ST}^2}\right)^{p_\mathrm{ST}}\right]\frac{\delta_\mathrm{ST}}{\sigma_{M_\mathrm{h}}}\times\nonumber\\
    &\times\exp\left(-\frac{a_\mathrm{ST}\delta_\mathrm{ST}^2}{2\sigma_{M_\mathrm{h}}^2}\right)\ ,
\end{align}
where $\overline{\rho}_\mathrm{m}$ is the average comoving matter energy density, $\sigma_{M_\mathrm{h}}^2$ is the variance of the density field smoothed over a mass scale $M_\mathrm{h}$, $A_\mathrm{ST} = 0.3222$, $a_\mathrm{ST} = 0.707$, $p_\mathrm{ST} = 0.3$ and $\delta_\mathrm{ST} = 1.686$. The mass variance is defined as:
\begin{align}
    \label{eq:sigmasq_M}
    \sigma^2_{M_\mathrm{h}} &= \int\frac{d^3k}{(2\pi)^3}W_{M_\mathrm{h}}^2(k)T^2_\zeta(k,z)P_\zeta(k)\ ,
\end{align}
with $W_{M_\mathrm{h}}$ a window function, and $T_\zeta$ and $P_\zeta$ the transfer function and primordial power spectrum of the comoving curvature perturbation $\zeta$, respectively. Unless otherwise stated, we use for the window function a spherical top hat in real space, which in Fourier space reads:
\begin{align}
    W_{M_\mathrm{h}}(k) =& \frac{3\sin\left(kR\right) - 3kR\cos\left(kR\right)}{\left(kR\right)^3}\ ,
\end{align}
where $R(M_\mathrm{h}) = [3M_\mathrm{h}/(4\pi\overline{\rho}_\mathrm{m})]^{1/3}$ is the Lagrangian radius (also known as the filter scale). This HMF has been tested against N-body simulations at the redshifts of interest~\cite{Lukic:2007fc, Schneider:2014rda}, including in a specific study of UV LFs~\cite{Tacchella:2018qny}.

As for the halo-galaxy connection, we use three astrophysical models to translate $M_\mathrm{h}$ into $M_\mathrm{UV}$, with differing assumptions about halo accretion, star-to-halo-mass ratios, and UV emission. These are detailed in our companion paper~\cite{Sabti:2021xvh}, where they are shown to produce consistent cosmological results. Here we will simply summarise our fiducial model. 
The star-formation rate (SFR) $\dot{M}_*$ of high-redshift galaxies strongly depends on their host-halo mass~\cite{Moster_2018, Wechsler:2018pic, Behroozi:2019kql}. In particular, the SFR is expected to peak for galaxies hosted in halos similar to that of the Milky Way, and to decrease for both smaller and bigger galaxies~\cite{Sun_2016}. This is due to a variety of different baryonic feedback processes, such as active galactic nuclei, supernovae shocks, and stellar winds~\cite{Fabian:2012xr, Kay:2001hq}. We model this behaviour by assuming a double-power law relation between the SFR of a halo and its accretion rate:
\begin{align}
    \label{eq:ftilde}
    \widetilde{f_*} = \dfrac{\dot{M}_*}{\dot{M}_\mathrm{h}} = \dfrac{\epsilon_*}{\left(\dfrac{M_\mathrm{h}}{M_c}\right)^{\alpha_*}+\left(\dfrac{M_\mathrm{h}}{M_c}\right)^{\beta_*}}\ ,
\end{align}
where $\alpha_* \leq 0$, $\beta_* \geq 0$, $\epsilon_*\geq 0$ and $M_c \geq 0$ are all free parameters, which control the slope of the faint end, slope of the bright end, the star-formation efficiency and mass at which the SFR peaks, respectively. In our fiducial model, we keep $\alpha_*$ and $\beta_*$ independent of redshift, whereas we allow $\epsilon_*$ and $M_c$ to evolve as power-laws of $z$ (see~\cite{Sabti:2021xvh} for more details). The SFR and UV luminosity $L_\mathrm{UV}$ of a galaxy are related as~\cite{Madau:1997pg, Kennicutt:1998zb}
\begin{align}
    \label{eq:Mstardot_LUV}
    \dot{M}_* = \kappa_\mathrm{UV}L_\mathrm{UV}\ ,
\end{align}
where $\kappa_\mathrm{UV} = 1.15\times 10^{-28}\, M_\odot\mathrm{\,s\, erg}^{-1}\mathrm{yr}^{-1}$ is a conversion factor obtained from stellar-population-synthesis models\footnote{This parameter could be varied in our analysis, though it is fully degenerate with $\epsilon_*$.}~\cite{Madau:2014bja}. The UV luminosity can be expressed in terms of the absolute magnitude through~\cite{Oke:1983nt}
\begin{align}
    \label{eq:LUV}
    \log_{10}\left(\frac{L_\mathrm{UV}}{\mathrm{erg \, s^{-1}}}\right) = 0.4\, (51.63 - M_\mathrm{UV})\ .
\end{align}
As a final step, we require an expression for the halo accretion rate $\dot M_h$. For this purpose, we turn to the extended Press-Schechter formalism, which provides a semi-analytical description for $\dot{M}_\mathrm{h}$ that agrees very well with the output of N-body simulations~\cite{Correa:2014xma}. Within this formalism, the halo mass grows exponentially during matter domination, and follows a power-law behaviour with $z$ at lower redshifts due to dark energy domination. The accretion rate is given by~\cite{Neistein:2006ak,Correa:2014xma}:
\begin{align}
    \label{eq:Mhdot_EPS}
    \dot{M}_\mathrm{h} = -\sqrt{\frac{2}{\pi}}\frac{(1+z)H(z)M_\mathrm{h}}{\sqrt{\sigma_{M_\mathrm{h}}^2(Q) - \sigma_{M_\mathrm{h}}^2}}\frac{1.686}{D^2(z)}\frac{\mathrm{d}D(z)}{\mathrm{d}z}\ .
\end{align}
In this equation, $D(z)$ is the linear growth factor and $\sigma_{M_\mathrm{h}}^2(Q) \equiv \sigma^2(M_\mathrm{h}/Q)$ is the rescaled mass variance, where $Q$ is a free parameter. This latter quantity is calibrated with N-body simulations, and we allow it to vary within the range $Q=1.5-2.5$ found by previous works~\cite{Neistein:2006ak,Schneider:2020xmf}. 

\begin{figure}[t!]
    \centering
    \includegraphics[width=\linewidth]{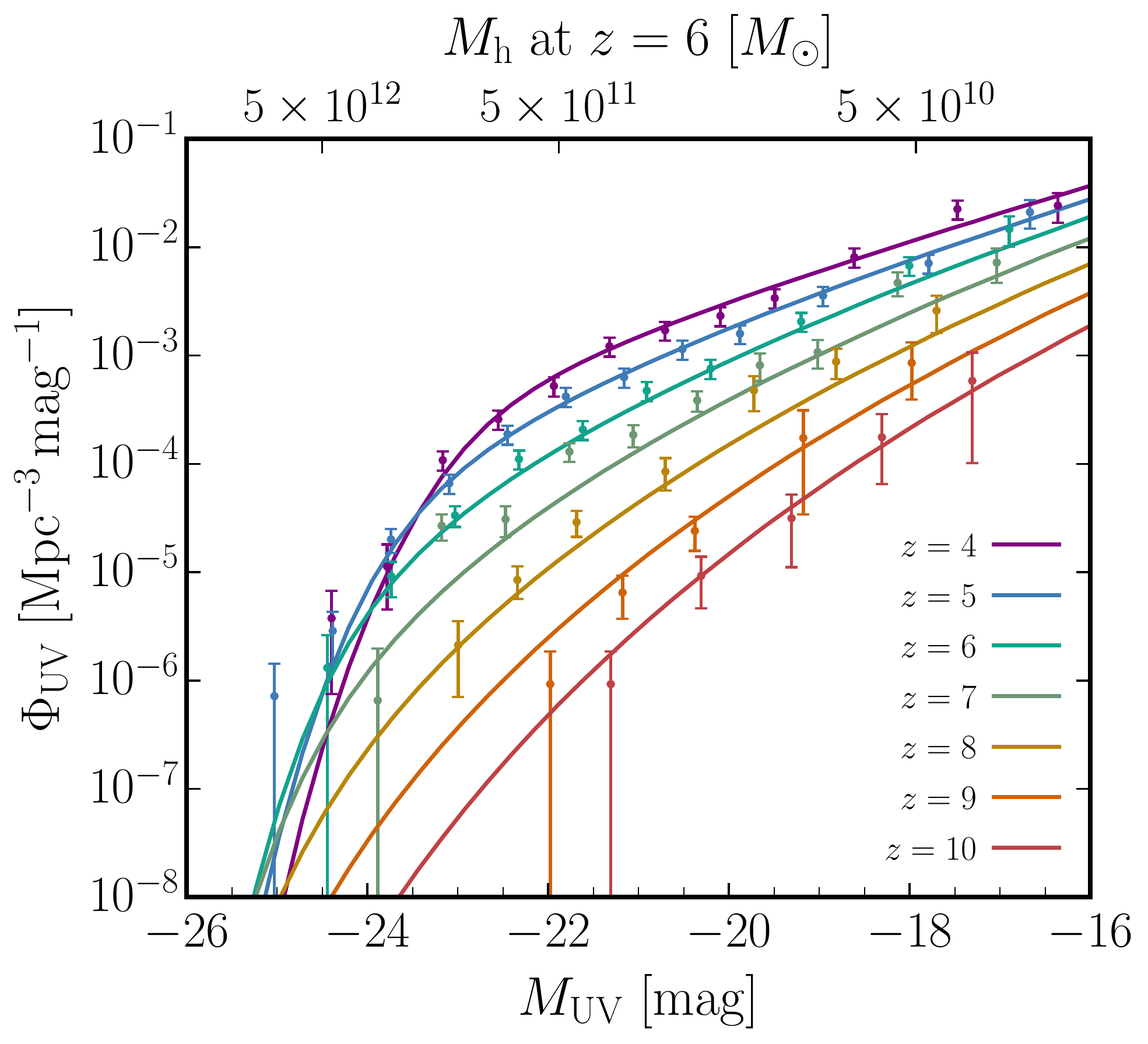}
    \caption{Global fit of our fiducial model to the UV LF data from the HST~\cite{Oesch_2018, Bouwens_2021}. The data points shown here are corrected for dust attenuation and the Alcock-Paczy\'{n}ski effect. For illustration purposes, the top axis shows the corresponding halo masses at redshift $z = 6$ within our best-fit model.
    }
    \label{fig:v2_bestfit}
\end{figure}

In summary, we use Eqs.~\eqref{eq:ftilde} and~\eqref{eq:Mhdot_EPS} to obtain the SFR as a function of halo mass $M_\mathrm{h}$, including scatter as described in~\cite{Sabti:2021xvh}. The SFR is then expressed in terms of a UV magnitude using Eqs.~\eqref{eq:Mstardot_LUV} and~\eqref{eq:LUV}. From this, we can straightforwardly compute the UV luminosity function in Eq.~\eqref{eq:UVLFdef}. We show our best-fit model in Fig.~\ref{fig:v2_bestfit}, which is in good agreement with the HST data.\\

\begin{figure}[t!]
    \centering
    \includegraphics[width=\linewidth]{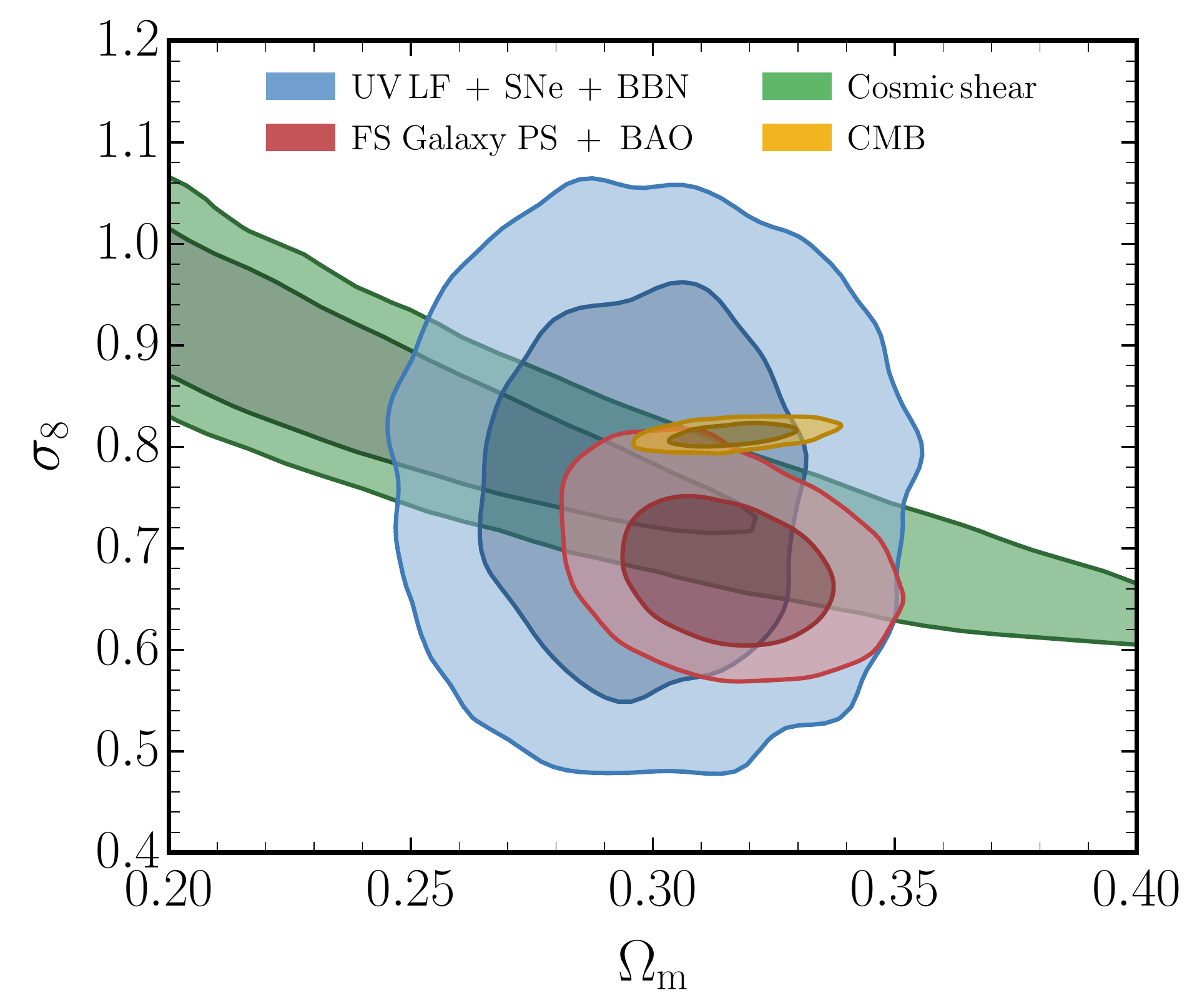}
    \caption{Posteriors for the large-scale clustering amplitude $\sigma_8$ versus the matter density parameter $\Omega_\mathrm{m}$. The inner (outer) contours represent the 68\% (95\%) confidence levels. 
    Our result (in blue, where we used HST UV LF data~\cite{Oesch_2018, Bouwens_2021}, Pantheon supernovae distance moduli determinations~\cite{Scolnic:2017caz, Jones:2017udy}, and baryon density inferences from primordial-abundance measurements~\cite{Pisanti:2020efz})
    is shown along the KV450+DES-Y1 cosmic-shear analysis from~\cite{Joudaki:2019pmv} (green), the BOSS full-shape galaxy-power-spectra and BAO analysis from~\cite{Philcox:2020vvt} (red), and the Planck CMB analysis from~\cite{Aghanim:2018eyx} (yellow). Note that the supernovae and primordial-abundance data used in our analysis here only constrain $\Omega_\mathrm{m}$ and $\omega_\mathrm{b}$, respectively, and do not contain any information on $\sigma_8$. In all cases, the total sum of neutrino masses is fixed to $\sum m_\nu = 0.06\,\mathrm{eV}$.
    }
    \label{fig:sigma8_Omegam}
\end{figure}

\textit{Clustering Amplitude.\ ---}
The raw UV LF data roughly covers absolute magnitudes from $M_{\rm UV}=-23$ to $-16$, which correspond to {\it typical} halo masses of $M_\mathrm{h}\sim 10^{10}-10^{12}\, \Msun$ at redshift $z = 6$ (assuming our fiducial model, see also Fig.~\ref{fig:v2_bestfit}). While these halos are rather common today, at high redshifts the high-mass (bright) end of the UV LF is in the exponential tail of the halo mass function, and thus particularly sensitive to changes in the amplitude of clustering. We will now quantify what clustering information can be extracted from the UV LFs. A full description of our analysis pipeline is provided in our companion paper~\cite{Sabti:2021xvh}. 

We start by measuring the large-scale clustering amplitude $\sigma_8 \equiv \sigma(R=8/h\,\mathrm{Mpc})$, which allows us to contextualise our results with other clustering measurements, both at lower and higher redshifts. We show our posteriors for $\sigma_8$ and $\Omega_\mathrm{m}$ in Fig.~\ref{fig:sigma8_Omegam}, along with those from CMB, cosmic shear, and galaxy-clustering observations. After marginalising over all cosmological and astrophysical parameters, we obtain a measurement of $\sigma_8$ that reads:
\begin{align}
    \label{eq:sigma8_fiducial}
    \sigma_8 = 0.76^{+0.12}_{-0.14}\ ,
\end{align}
at 68\% CL. 
We thus find that UV LFs can measure $\sigma_8$ within ${\sim}15\%$ uncertainty, a factor of few less constraining than what is obtained with the other data sets, though independent of them (see also~\cite{Sahlen:2021bqt} for a recent study).

A key advantage of the UV LFs is their ability to probe smaller halo masses, where deviations from CDM may first appear. Thus, we now turn to measuring the amplitude of matter fluctuations at small scales. Rather than focusing on any particular non-CDM model, we will obtain model-agnostic measurements of the matter power spectrum at large wavenumbers $k$, which can then be applied to a wide range of models. We will follow a simple approach in doing so: we divide the power spectrum into four bins, whose amplitudes $a_{\mathrm{s},i}$ we will vary: 
\begin{align}
    \label{eq:Pk_bins}
    P(k) = 
    \begin{cases} P_k^{\Lambda\mathrm{CDM}} \qquad &\mathrm{if\ } k <  0.5\,\mathrm{Mpc}^{-1} \\
    a_\mathrm{s,2}P_k^{\Lambda\mathrm{CDM}} \qquad &\mathrm{if\ } 0.5 \leq k <  2.25\,\mathrm{Mpc}^{-1} \\
    a_\mathrm{s,3}P_k^{\Lambda\mathrm{CDM}} \qquad &\mathrm{if\ } 2.25 \leq k <  10\,\mathrm{Mpc}^{-1} \\
    a_{\mathrm{s,4}}P_k^{\Lambda\mathrm{CDM}} \qquad &\mathrm{if\ } k \geq  10\,\mathrm{Mpc}^{-1}
    \end{cases}\ , 
\end{align}
where $P_k^{\Lambda\mathrm{CDM}}$ is the matter power spectrum in $\Lambda$CDM. We emphasise that the amplitudes $a_{\mathrm{s},i}$ are \emph{relative} to the overall scalar amplitude $A_{\rm s}$, which is the amplitude of $P_k^{\Lambda\mathrm{CDM}}$. We use Planck 2018 TTTEEE+lowE+lensing data~\cite{Planck:2019nip} to constrain the large-scale behaviour of the power spectrum (and thus $A_{\rm s}$, which also acts as the amplitude of the first bin). The bins have been chosen so that CMB data can mainly probe the first one ($k <  0.5\, \mathrm{Mpc}^{-1}$, see e.g.~\cite{Chabanier:2019eai}), whereas the last one at very small scales ($k > 10\,\mathrm{Mpc}^{-1} $) will not be well measured even by the UV LFs. We divide the intermediate range $0.5\, \Mpcinv < k < 10\,\Mpcinv$ into two bins, whose amplitudes we can measure with the UV LFs\footnote{Using more bins would weaken our constraints, since the amplitudes of the bins are strongly correlated with each other. Two bins is therefore a reasonable compromise between obtaining strong constraints and resolving the $k$-behaviour.}. 
The amplitudes $a_{\mathrm s,i}$ (with fiducial values of unity) are varied independently from $10^{-9}$ to $10^{9}$, assuming a log-flat prior. 

Since we essentially allow for a cut-off in the matter power spectrum, we need to be careful with our choice of the window function in Eq.~\eqref{eq:sigmasq_M}. Using a real-space top-hat window function with a truncated power spectrum can lead to $\sigma_{M_\mathrm{h}}$ to keep increasing even for masses below the cut-off scale~\cite{Schneider:2014rda, Benson:2012su}. This is a well-known issue and can be circumvented by using a sharp-$k$ window function~\cite{Bertschinger:2006nq, Schneider:2013ria}:
\begin{align}
    W_{M_\mathrm{h}}(k) = \Theta(1-kR)\ ,
\end{align}
where $\Theta$ is the Heaviside step function. A side effect of using the sharp-$k$ filter is that the relation between halo mass and filter scale $R$ is not well defined. We follow~\cite{Schneider:2014rda, Schneider:2018xba} and introduce a new parameter $c=2.5$ in the definition of the halo mass $M_\mathrm{h} = 4\pi\overline{\rho}_\mathrm{m}(cR)^3/3$, which is found to fit well their simulations with cold, warm and fuzzy DM. Following these same references, one would have to set $a_\mathrm{ST}$ in Eq.~\eqref{eq:f_ST} to unity, as the new variable $c$ takes over its role in calibrating the HMF. We conservatively allow for additional freedom in the HMF by varying $a_\mathrm{ST}$ between 0.9 and 1, which ensures that the CDM mass function is always included as a prediction in our model.

\begin{figure}[t!]
    \centering
    \includegraphics[width=\linewidth]{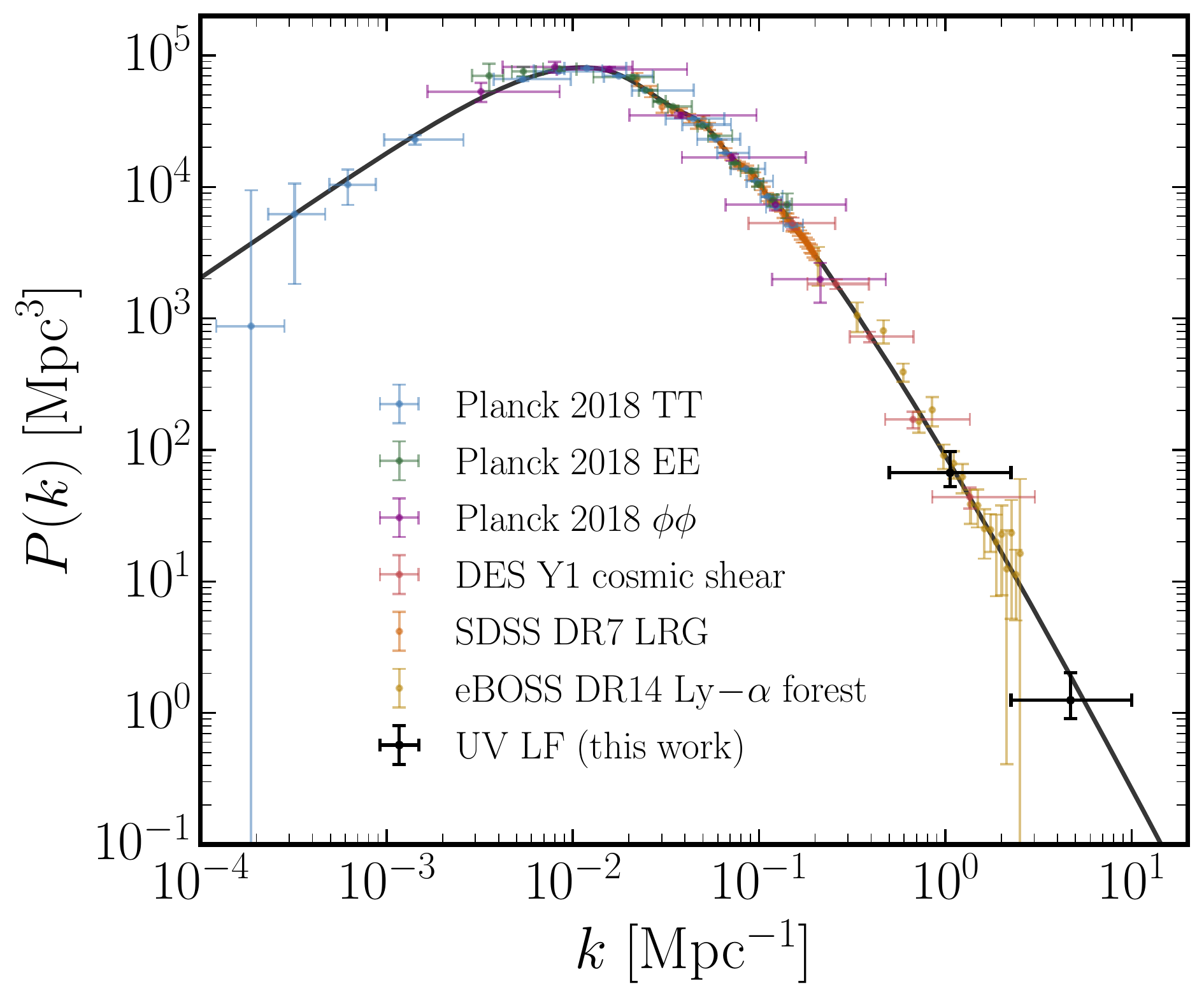}
    \caption{Measurements of the matter power spectrum (of DM and baryons only) linearly extrapolated to redshift $z=0$. The two black data points are the results of this work and are obtained with the UV LF data from~\cite{Oesch_2018, Bouwens_2021}, where we imposed a prior from Planck 2018 CMB observations (TTTEEE+lowE+lensing) to constrain the small-$k$ behaviour. The coloured data points represent measurements using Planck 2018 CMB~\cite{Aghanim:2018eyx}, DES cosmic shear~\cite{DES:2017qwj}, SDSS galaxy clustering~\cite{Reid:2009xm} and SDSS Lyman-$\alpha$~\cite{SDSS:2017bih} data (see~\cite{Chabanier:2019eai} for more details). The black line is the prediction within $\Lambda$CDM, using the best-fit values from Planck 2018~\cite{Aghanim:2018eyx}. All uncertainties in this figure are at $68\%$ CL.
    }
    \label{fig:mpk} 
\end{figure}

We use our fiducial UV LF model and run our analysis with the 2021 HST data compiled in~\cite{Bouwens_2021}, in conjunction with the Planck 2018 TTTEEE+lowE+lensing data~\cite{Planck:2019nip}. We present our main results in Fig.~\ref{fig:mpk}. This figure displays our measurement of the small-scale matter power spectrum alongside data points from a number of other probes~\cite{Chabanier:2019eai}. The black data points are obtained by computing the power spectrum at the centre of each bin (as in Eq.~\eqref{eq:Pk_bins}), and marginalising over all cosmological and astrophysical parameters. We measure the small-scale amplitudes (relative to $\Lambda$CDM) in Eq.~\eqref{eq:Pk_bins} to be $a_{\mathrm{s},2} = 0.93_{-0.25}^{+0.34}$ and $a_{\mathrm{s},3} = 0.66_{-0.17}^{+0.43}$ at 68\% CL. Our UV LF analysis is able to reach smaller scales than other current cosmological probes, which provides a new lamppost to understand the clustering properties of dark matter~\cite{Banik:2019smi, Jethwa:2016gra, DES:2020fxi, Newton:2020cog, Hsueh:2019ynk, Gilman:2019nap, Enzi:2020ieg, Nadler:2021dft, Nadler:2019zrb}. We find that the matter power spectrum is consistent with the theoretical prediction of a standard $\Lambda$CDM cosmology up to $k=10\,\Mpcinv$, which disfavours alternatives that suppress power at these scales, such as warm~\cite{Bode:2000gq, Boyarsky:2018tvu} or fuzzy~\cite{Marsh:2015xka, Hui:2021tkt} DM. \\

\textit{Conclusions.\ ---} UV galaxy luminosity functions capture a wealth of information about the Universe around the epoch of cosmic reionisation. In addition to shedding light on the astrophysics of this interesting era, we have shown that the same data can be used to measure the clustering of matter at smaller scales and higher redshifts than currently accessible. In particular, here we used UV LF data from observations of the Hubble Space Telescope~\cite{Oesch_2018, Bouwens_2021} to derive new constraints on the matter power spectrum at wavenumbers $k=0.5-10\,\Mpcinv$ and redshifts $z=4-10$. In this range, dark-matter halo collapse is still not in the deep non-linear regime (i.e., the satellite fraction is negligible) and the dust attenuation only affects the brightest galaxies, which simplifies the modelling. Throughout the text, we focused on our fiducial model for the halo-galaxy connection. As a cross-check, we have performed the same study using the two other astrophysical models detailed in our companion paper~\cite{Sabti:2021xvh}, and found good agreement among the three models, due to our marginalisation over the astrophysical parameters. In addition, we have used an alternative determination of the UV LFs from~\cite{Finkelstein_2015} and found consistent results for $\sigma_8$ and the amplitudes in Eq.~\eqref{eq:Pk_bins} within roughly $1\sigma$ and $2\sigma$, respectively. Finally, we find that using the Reed mass function or different calibrations for the dust extinction does not alter our conclusions significantly, as in both cases mainly the bright end of the UV LF (where Poisson errors are already large) is affected, see~\cite{Sabti:2021xvh}. We note that if a cut-off in the UV LF were detected, one would have to proceed with caution. For example, one could do a careful model comparison (e.g., with a Bayesian method) to determine whether the data prefers a dark-matter cut-off over an astrophysical one, given their different shapes.

Our analysis here establishes the UV LF as a powerful cosmic probe of small-scale structure, providing us with valuable insights beyond the frameworks of specific dark-matter or inflationary models~\cite{Bozek:2014uqa, Schultz:2014eia, Dayal:2014nva, Corasaniti:2016epp, Menci:2017nsr,Menci:2018lis, Rudakovskyi:2021jyf,Chevallard:2014sxa, Yoshiura:2020soa,Sabti:2020ser}. Together with current large-scale cosmological data sets, the UV LF expands our knowledge on the clustering of matter to cover nearly five orders of magnitude in scales ($10^{-4}\,\mathrm{Mpc}^{-1} < k < 10\,\mathrm{Mpc}^{-1}$, see Fig.~\ref{fig:mpk}). In the near future, the James Webb Space Telescope~\cite{Gardner:2006ky} and Nancy Gracy Roman Space Telescope~\cite{Spergel:2015sza} will not only observe galaxies at higher redshifts than covered by current HST data, but also probe halos with smaller masses. This provides us with an exciting outlook on the study of the growth and clustering of matter.\\

\textit{Acknowledgements.\ ---}  We thank Mikhail Ivanov for providing us MCMC chains of the BOSS FS + BAO analysis, and Marius Millea for a helping hand in extracting data points of the matter power spectrum from complementary probes. NS is a recipient of a King's College London NMS Faculty Studentship. JBM is supported by a Clay fellowship at the Smithsonian Astrophysical Observatory. IFAE is partially funded by the CERCA program of the Generalitat de Catalunya. The research leading to these results has received funding from the Spanish Ministry of Science and Innovation (PID2020-115845GB-I00/AEI/10.13039/501100011033). DB is supported by a `Ayuda Beatriz Galindo Senior' from the Spanish `Ministerio de Universidades', grant BG20/00228. We acknowledge the use of the public cosmological codes \texttt{CLASS}~\cite{Blas:2011rf,Lesgourgues:2011re} and \texttt{MontePython}~\cite{Audren:2012wb, Brinckmann:2018cvx}. The simulations in this work were performed on the Rosalind research computing facility at King’s College London, and the FASRC Cannon cluster supported by the FAS Division of Science Research Computing Group at Harvard University.

\bibliographystyle{apsrev4-1}
\bibliography{biblio}

\end{document}